\documentclass[twocolumn,showpacs,preprintnumbers,amsmath,
amssymb]{revtex4}
\usepackage{graphicx}
\usepackage{dcolumn}
\usepackage{bm}
\usepackage{tensor}
\begin{document}
 
\title{Rotational Doppler shift of the phase-conjugated 
photon.}
 
\author{A.Yu.Okulov}
\email{alexey.okulov@gmail.com}
\homepage{http://okulov-official.narod.ru }
\affiliation{Russian Academy of Sciences, 
Moscow, Russia}

\date{\ April 12, 2011}

\begin{abstract}

{ The rotational Doppler shift of a photon 
with orbital angular momentum $\pm \ell \hbar$ is shown 
to be an even multiple of the angular frequency $\Omega$ 
of the reference 
frame rotation when photon is 
reflected from the phase-conjugating mirror. 
The one-arm phase-conjugating 
interferometer is considered. It contains $N$ Dove prisms 
or other angular momentum altering elements rotating in opposite 
directions. 
When such interferometer is placed in the 
rotating vehicle the 
$\delta \omega=4 (N+1/2) \ell \cdot \Omega$ 
rotational Doppler shift appears and rotation of the 
helical interference pattern with angular 
frequency $\delta \omega /{2 \ell}$ occurs.  
The accumulation of angular Doppler shift via successive 
passages through the $N$ image-inverting prisms is due 
to the phase conjugation, for conventional parabolic 
retroreflector the accumulation is absent. 
The features of such a vortex phase 
conjugating interferometry 
at the single photon level are discussed.}

\end{abstract}

\maketitle

\vspace{1cm} 
\section {Introduction.}
Single photon interferometry utilizes the superposition 
of a mutually coherent (phase-locked) quantum 
states $\Psi_j$ \cite {Scully:1997} to whom photon 
belongs simultaneously. The interference 
pattern depends on a method of $\Psi_j$ preparation. 
The double-slit Young interferometer creates two 
free-space wavefunctions $\Psi_1$,$\Psi_2$, whose interference pattern 
produced by detection of the individual photons is recorded 
by an array of detectors or a photographic plate located 
in the near or far field. In Mach-Zehnder 
configuration \cite {Barnett:2002,Soskin:2008} 
two wavefunctions 
separated by entrance beamsplitter recombine at the output 
beamsplitter. 
The Michelson interferometer recombines at the 
input beamsplitter 
two $retroflected$ quantum states provided these states  
are phase-locked 
and their path difference $\delta L$ is smaller 
than the coherence length $L_{c}$. 
Thus interference pattern is simply 
$\sim [1+{\it V(\delta L)}\cdot \cos(2 k \cdot \delta L)]$, 
where $V(\delta L)$ is a visibility or second-order 
correlation function and 
$k=2 \pi / \lambda$. 
When retroflection is accompanied by wavefront reversal 
(PC) realized with phase-conjugating mirrors (PCM)  
\cite {Boyd:2007} based upon Stimulated Brillouin 
scattering \cite {Zeldovich:1985,Basov:1980}, 
photorefractivity \cite {Woerdemann:2009,Mamaev:1996} or 
holographic PCM's, 
the optical path $\delta L$ 
difference is almost entirely compensated due to PC.
Noteworthy the small phase lag 
due to the relatively small frequency 
shift $\delta \omega =\omega_f - \omega_b$ 
arising due to the excitation 
of internal waves inside PCM 
volume \cite {Okulov:2008}, where $\omega_f$ 
and $\omega_b$ are the carrier 
frequencies of incident 
and PC-reflected photon respectively. This 
leads to the interference term 
$1+{\it V}(\delta L)\cdot \cos(\delta k \cdot \delta L)$, 
where $\delta k=\delta \omega /c$ \cite {Basov:1980}.

We study the photon in the optical vortex quantum 
state \cite {Barnett:2002,Soskin:2008} with topological 
charge $\ell$, where the angular 
momentum $L_z=\pm \ell \cdot \hbar$ is due to the 
phase singularity located at propagation axis $Z$ (hereafter 
the spin component of angular momentum \cite {Beth:1936} 
is supposed to be zero due to the linear polarization). 
It is convinient to use 
the single-photon wavefunctions which coincide  
with the positive frequency component of the electric field 
envelope 
$ {\bf |\Psi}>= {\sqrt {2\epsilon_0}} \cdot { E(t,\vec r)}$ 
\cite {Sipe:1995}. The square modulus of $\Psi$ 
is proportional to the energy density of 
the $continuous$ $wave$ 
laser beams (CW) 
and to the photons count rate in 
a different fringes of the interference pattern for the 
single-photon experiments \cite{Kapon:2004}. We will 
assume $\Psi$ to have the form of the 
Laguerre-Gaussian beam (LG) with $\ell \hbar$ 
orbital angular momentum (OAM) per 
photon \cite {Okulov:2008} 
but any other isolated 
vortex solutions, e.g. Bessel vortices  
\cite {Eberly:1987,Sepulveda:2009} will demonstrate 
the same final results: 
\begin{eqnarray}
\label{pump1}
{{\bf \Psi}_{(f,b)}(z,r,\theta,t)}
\sim \sqrt{2\epsilon_0} \cdot
{\frac {  \exp [ i( -\omega_{(f,b)}t 
\pm k_{(f,b)} z) \pm i{\ell}\theta ]} 
{ {(1{+}iz/z_R)}} } 
&& \nonumber \\
{{\ E}_{(f,b)}}{({r}/{D_0})^{|\ell|}}
\exp  [  - 
{\frac {r^2}{{D_0}^2(1{+}iz/z_R)}} ], 
z_R=k_{(f,b)} {D_0}^2{\:} {\:} 
\end{eqnarray}
where the cylindrical coordinates $(z,r,\theta)$ 
are used, $D_0$ is the vortex radius, $z_R$ 
is Rayleigh range, $\Psi_f, E_f$ stands for the forward 
wave, propagating in positive Z-direction, 
$\Psi_b, E_b$ stands for the wave, propagating in 
the negative one. Of special interest is the 
sub-$Hz$ - order angular frequency splitting 
$\delta \omega=c (k_f - k_b) $ 
which appears due to the slow mechanical rotation 
of the setup 
\cite {Soskin:2008,Dholakia:2002}. 
It was already shown that rotation of the 
$\lambda/2$ waveplate 
with angular 
frequency $\Omega \sim 2\pi (1-100) rad/s$ in a one arm of the 
Mach-Zehnder interferometer induces the rotational 
Doppler shift (RDS) $\delta \omega=2 \Omega \ell$ for circularily 
polarized broadband CW with linewidth 
$\Delta \omega /2 \pi \simeq 10^{10} Hz$. In this 
configuration the broadband 
spectrum was shifted $as$ $a$ $whole$ 
via mechanical rotation (by angular 
Doppler effect) at $\delta \omega / 2 \pi=\pm 2 \cdot 7 Hz$ 
and the beats at  
the output mirror induced an appropriate 
rotation of the interference 
pattern \cite {Dholakia:2002D}. 
\section {Phase-conjugating mirror in a rest frame.}
\begin{figure} 
{\includegraphics
[width=0.99\linewidth , angle=0]{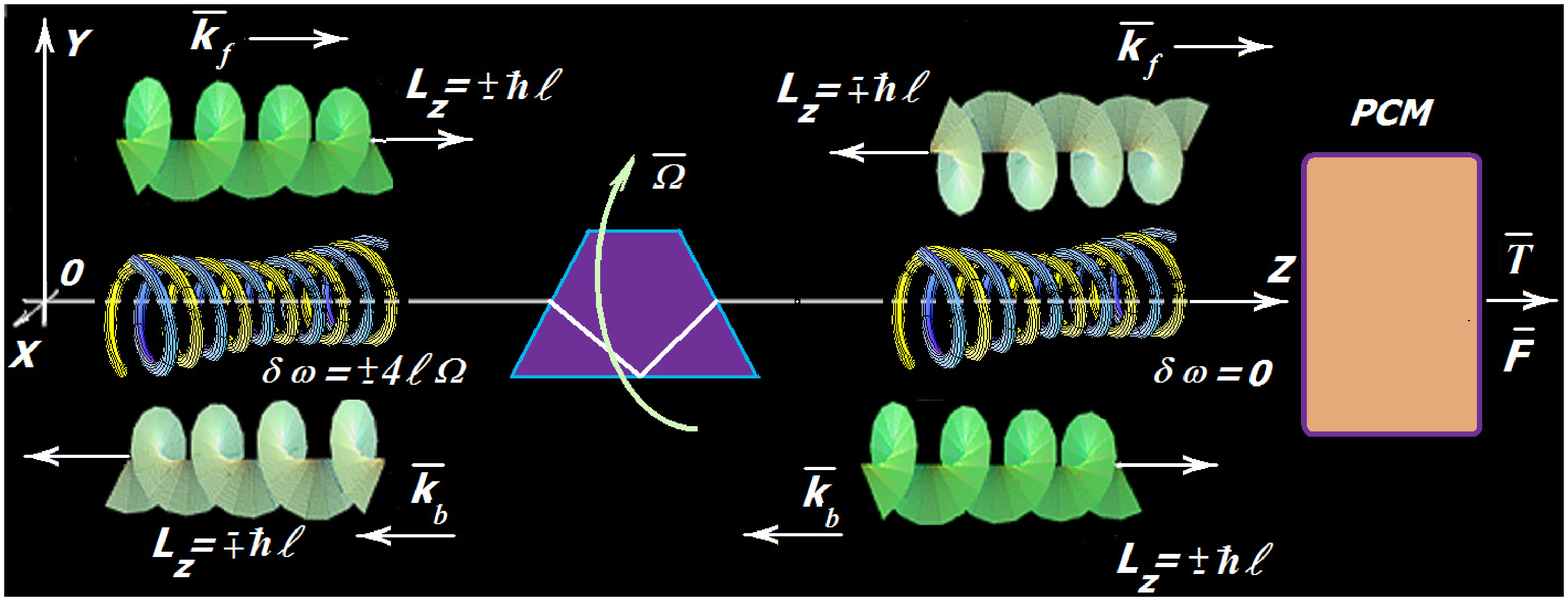}}
\caption{(Color online)Additivity of RDS 
for the PCM-reflected photon in the $rest$ $frame$. Rotation of the Dove prism (positive $\Omega$) 
decreases frequency to $-2 \ell \cdot \Omega$ of 
the co-rotating 
incident photon with $L_z=+ \ell \hbar$. Reflection from PCM 
alters $L_z$ projection to the opposite one and clockwise  
rotation of Dove prism (as seen to backward photon) 
again decreases 
the frequency of co-rotating photon 
to $-2 \ell \cdot \Omega$.  
Helical interference 
pattern is static between prism and PCM 
(where $\delta \omega = 0$) and 
rotates $before$ prism with angular 
velocity $\delta \omega =-2 \Omega$.}
\label{fig.1}
\end{figure}

Let us consider first the single-arm phase-conjugating vortex 
interferometer (PCVI) when PC-mirror is in the 
rest frame. Due to the 
reflection from PCM the helical photon with 
a $linear$ polarization proves 
to be in a superposition of the 
two counter propagating quantum 
states $\Psi_{f,b}$ (fig.\ref {fig.1}). 
Currently the best candidate 
for the $ideal$ single-photon PCM is a thick 
hologram written 
with sufficiently high diffraction efficiency 
($R \sim 0.9$) for the 
$\ell$ - charged optical 
vortex \cite {Soskin:2008,Volostnikov:1989,Felde:2008}. 
In such a case the amplitudes of 
forward and backward fields are close 
to each other 
and visibility of the interference pattern $V(\delta L)$ 
is close to 1, provided that 
coherence length $L_c \sim 2\pi c/ \Delta \omega$ 
is bigger than the doubled length of PCVI.
 
The $ideal$ PCM ensures the perfect coincidence of the  
helical phase surfaces 
of the counter propagating optical vortices $\Psi_{f,b}$ 
and zeros of their electric field amplitudes on 
$Z$ axis. 
In contrast to the speckle fields 
whose interference pattern 
is composed of intertwined Archimedean 
screws \cite {Okulov:2009} in PCVI the 
isolated Archimedean 
screw pattern appears both for the $single$ photon with LG 
wavefunction and for CW resulting in the intensity profile $I_{tw}$ 
composed of $2 \ell$ twisted $fringes$ 
\cite {Woerdemann:2009,Bhattacharya:2007,Okulov:2010josa}:
\begin{eqnarray}
\label{Archimedean_inter_patt}
z'=z-z_{pc},{|\Psi |^2}={|\Psi_f + \Psi_b |^2}\sim
{I_{_{tw}}}{(z',r,\theta,t )} =
{\:}{\:}{\:}{\:}{\:}{\:}{\:}{\:}{\:}{\:}
&& \nonumber \\
{\frac {2 {\epsilon_{_0} c}|{{E}_{(f,b)}}|^2 
{2^{(|\ell| +1)}}{({r}/{D_0})^{2|\ell|}}}
{\pi {\ell !}{D_0}^2 { (1+{z'}^2/{z_R}^2)  }}}\cdot
{\exp[-{\frac {2 r^2}{{D_0}^2(1+{z'}^2/{z_R}^2)}}]}
&& \nonumber \\
{[ 1 +R^2+ {\:}2R \cdot 
\cos[\delta \omega \cdot t - (k_f+k_b) z' + 
{2\ell}{\:}\theta ]]}, {\:}{\:}{\:}{\:}{\:}{\:}
\end{eqnarray}
where $z_{pc}$ is location of PCM entrance window.
The angular speed of pattern rotation 
$\dot \theta=\delta \omega /2 \ell$ is given 
by the differentiation 
of the self similar argument $2 \theta(t) \cdot \ell + 
\delta \omega \cdot t - (k_f + k_b)z'{\:}{\:}$\cite {Dholakia:2002D} vs 
time $t$.
Consider the origin of RDS $\delta \omega $
\cite {Garetz:1981,Birula:1997,Soskin:2008,Padgett:1998}
for the photon with topological 
charge $\ell$ after the double 
passage through a Dove prism rotating 
with angular 
velocity $\vec \Omega$ and reflection from PCM.
OAM projection on propagation axis  
$Z$ is $<\Psi_{f,b}|\hat L_z|\Psi_{f,b}>=\pm \ell \hbar$, where 
$\hat L_z= - i\hbar \cdot \partial / \partial  \theta$. 

The RDS occurs because the optical torque on a 
slowly rotating 
element changes the angular momentum of the 
prism \cite {Okulov:2008,Beth:1936}. In its 
turn this changes 
the prism's angular velocity $\vec \Omega$ and such a change 
requires the 
energy supply. Because typical optical 
elements including prisms are 
macroscopic classical objects having 
the $continuous$ $spectrum$ of energies, in such case 
the energy 
$\hbar \omega_{f,b}$ hence 
the frequency of the photon may be changed continuously
\cite {Dholakia:2002D}. 

Noteworthy that without PCM the rotation of Dove prism with respect 
to the other components
of the optical setup would be highly 
sensitive to misalignments and this would require 
a very accurate tuning \cite {Soskin:2008}. The phase conjugation 
facilitates the adjustments and provides interference pattern 
with good visibility \cite {Basov:1980}.  
\begin{figure}
{\includegraphics
[width=0.99\linewidth , angle=0]{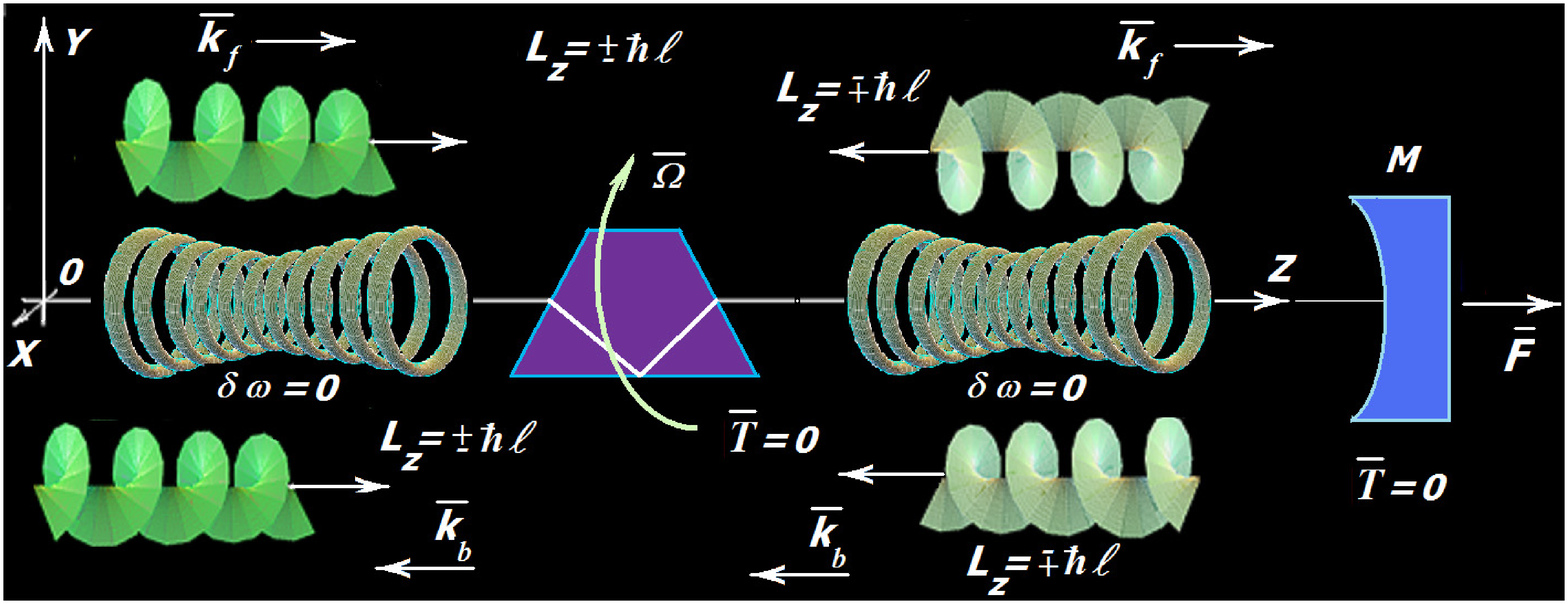}}
\caption{(Color online)Mutual cancellation of RDS 
for retroreflected photon. 
Rotation of the Dove prism 
again decreases frequency 
of the co-rotating 
forward photon with $L_z=+ \ell \hbar$ by 
$-2 \ell \cdot \Omega$. In 
backward propagation 
the Dove prism is counter rotating with respect to 
photon. Backward RDS is positive thus 
resulting $\delta \omega$ is zero hence 
toroidal interference 
pattern is static for all $Z$.}
\label{fig.2}
\end{figure}

In the following phase-conjugating optical 
interferometer the photon's OAM direction is 
altered as well \cite {Pepper:1980, Okulov:2008} (fig.\ref{fig.1}). 
Let the optical 
vortex  $E_{f}(t,\vec r)$ of charge $\ell$ to pass 
through a rotating Dove prism and to be reflected with 
 $E_{b}(t,\vec r)$ from 
a some $ideal$ (PCM). 
The non-rotating PCM is supposed 
to produce $no$ frequency shift as it happens in some cases in 
photorefractive crystals \cite{Woerdemann:2009,Mamaev:1996}, 
degenerate four-wave mixing 
\cite {Pepper:1980,Zeldovich:1985}, 
and holographic PC couplers 
\cite {Soskin:2008,Volostnikov:1989}. 
Noteworthy that a small $10^{-1}-10 Hz$ 
frequency shifts in $BaTiO_3$ 
photorefractive PCM may mask the RDS. These 
additional frequency shifts due to slow internal 
charge waves and filamentation effects were reported 
in early 1980's yet \cite {Fisher:1988,Fischer:1986}.

Because space is homogeneous and isotropic 
the conservation of energy, momentum and angular momentum 
is expected \cite {Pitaevskii:1982}.
Reproducing the Dholakia's symmetry arguments 
\cite {Dholakia:2002D} adapted to the current case 
we have the following conservation laws for the 
angular momenta 
$L_z$ with respect to $z-axis$ and the energies 
of the incident photons and those  transmitted 
through the Dove prism, when the latter rotates 
with the angular velocity $\vec \Omega$: 
\begin{eqnarray} 
\label {ang_moment_conserv_photon} 
I_{zz} \cdot \Omega + L_z =
I_{zz} \cdot {\Omega}^{'} + L_z^{'}
&& \nonumber \\
\hbar \omega_f + \frac {I_{zz}\Omega^2}{2}=
\hbar \omega^{'} + \frac {I_{zz}{\Omega^{'}}^2}{2},
\end{eqnarray}
where $I_{zz}$ is the moment of inertia around $Z$-axis, 
left hand sides of this system correspond to the 
incident photon and the right hand sides correspond to the 
transmitted one. The co-rotation of the prism and the photon 
corresponds to the same sign of projections 
($\Omega, \Omega^{'}$) of $\vec \Omega$ and $L_z$ 
on z-axis before and after the photon's passage. 
For the incident $L_z=+\ell \hbar$ and 
passed $L_z^{'}=-\ell \hbar$
the eq.
(\ref{ang_moment_conserv_photon}) gives the difference 
of the angular velocities of the prism before and after 
the photon passage: 
\begin{equation}
\label {ang_velocity_change} 
\Omega - {\Omega}^{'} =- \frac {2 \ell \cdot \hbar}{I_{zz}}.
\end{equation}

This means that co-rotation increases the angular velocity 
of prism, because the energy is transmitted to prism by a virtue 
of the optical torque $|\vec T|=2 \ell \cdot P/{\omega_f}$, 
where $P$ is total power carried by LG 
\cite {Soskin:2008} hence Doppler frequency shift for the 
photon $\omega^{'}- \omega_f$ is negative: 
\begin{equation}
\label {freq_shift_photon} 
\delta \omega = \omega^{'}- \omega_f = \frac {I_{zz}}{2 \hbar}
(\Omega-\Omega^{'})(\Omega^{'}+\Omega)=
-2 \ell \cdot \Omega - \frac{2\ell \cdot \hbar}{I_{zz}}.
\end{equation}
Obviously in the counter rotating case, when 
projections of $\vec \Omega$ and $L_z$ are in the 
opposite directions, 
the rotational Doppler shift is positive. 
The net RDS during total forward-backward passage 
is additive due to PCM and this 
results in the net OAM 
change $\delta \omega = \pm 4 \ell \cdot \Omega$. 
The interference pattern and RDS will be the same for all 
PC-mirrors close to an ideal one, including just proposed 
$linear$ $loop$ PCM, which uses flat optical 
surfaces without any holographic 
element \cite {Okulov:2010josa}.
The frequency shift $\delta \omega$ is zero 
in between Dove prism and PCM and helical pattern is static there.
In the region before Dove prism, the frequency shift 
causes the clockwise ($\delta \omega= - 4 \ell \cdot \Omega$) or 
counterclockwise ($\delta \omega= + 4 \ell \cdot \Omega$) rotation 
of optical helix pattern \cite {Okulov:2008}. 
As stated before (\ref {Archimedean_inter_patt}) 
the angular speed of the 
optical helix rotation $\dot \theta = \pm 2 \Omega$ is 
smaller than $\delta \omega$.  

Because 
of the angular momentum conservation  
the PCM feels the $rotational$ $recoil$ which is proportional 
to topological 
charge $\ell$: $|\vec T_{pc}|= \ell \cdot P(\omega_f^{-1}+\omega_b^{-1})$. 
The Dove prism feels doubled 
torque : $|\vec T_{Dove}|= 2\ell \cdot P(\omega_f^{-1}+\omega_b^{-1})$. 
Note that Lebedev radiation pressure force 
is always directed 
in positive $Z$ direction and $|\vec F_{pc}|=2\cdot P/c$ 
is independent of $\ell$ \cite{Lebedev:1901}. 

In the absence of the truly phase-conjugating mirror 
when the forward beam 
is retroreflected by spherical mirror (fig.\ref{fig.2})  
without altering the 
angular momentum the interference pattern 
is a toroidal one $I_{tor}$ 
\cite{Rempe:2007,Okulov:2008,Woerdemann:2009,Sepulveda:2009}:
\begin{eqnarray}
\label{inter_patt4}
{|\Psi |^2}={|\Psi_f + \Psi_b |^2}\sim
{I_{_{tor}}}{(z',r,\theta,t )}=
{\:}{\:}{\:}{\:}{\:}{\:}{\:}{\:}{\:}{\:}
&& \nonumber \\
{\frac {2 {\epsilon_{_0} c}|{{E}_{(f,b)}}|^2 
{2^{(|\ell| +1)}}{({r}/{D_0})^{2|\ell|}}}
{\pi {\ell !}{D_0}^2 { (1+{z'}^2/{z_R}^2)  }}}\cdot
{\exp[-{\frac {2 r^2}{{D_0}^2(1+{z'}^2/{z_R}^2)}}]}
&& \nonumber \\
{[ 1 +R^2+ {\:}2R \cdot 
\cos[\delta \omega \cdot t-(k_f+k_b) z']]}
{\:}.{\:}{\:}{\:}{\:}
\end{eqnarray}
The RDS is not accumulated here (the Doppler shifts 
for the forward and backward photons cancel 
each other $\delta \omega=0$) 
because the truly phase-conjugation is absent 
hence toroidal interference pattern is static. 
The mechanical torques on prism 
$\vec T$ induced by OAM alternation will cancel each other too: 
$|\vec T|=2 \ell\cdot P(\omega_f^{-1}-\omega_b^{-1})\cong 0$.

\section {Phase-conjugating mirror in a rotating frame.}

The more fundamental 
case is a rotation of the all setup as a whole around a some axis 
and this general case is relevant to the detection of the slow rotations 
of the reference frame \cite {Scully:1997}. In contrast to Sagnac 
interferometer where frequency splitting and running 
interference pattern appears in active loop only the PCVI 
produces frequency splitting $\delta \omega$ even in the passive 
configuration (fig.\ref{fig.3}). 
Apparently the RDS $\delta \omega$ will be the same when CW laser is 
placed in both inertial (rest) frame and 
in the noninertial frame associated with rotating vehicle.
For the simplest case when the sole PCM rotates around 
propagation axis $Z$ of a twisted photon with charge $\ell$ 
the RDS appears due to the alternation of the photon's OAM. 
The eqs. (\ref{ang_moment_conserv_photon}) give again the 
frequency shift 
$\delta \omega = \omega_b - \omega_f$ due to the reflection from rotating PCM:

\begin{figure}
{\includegraphics
[width=1.0\linewidth , angle=0]{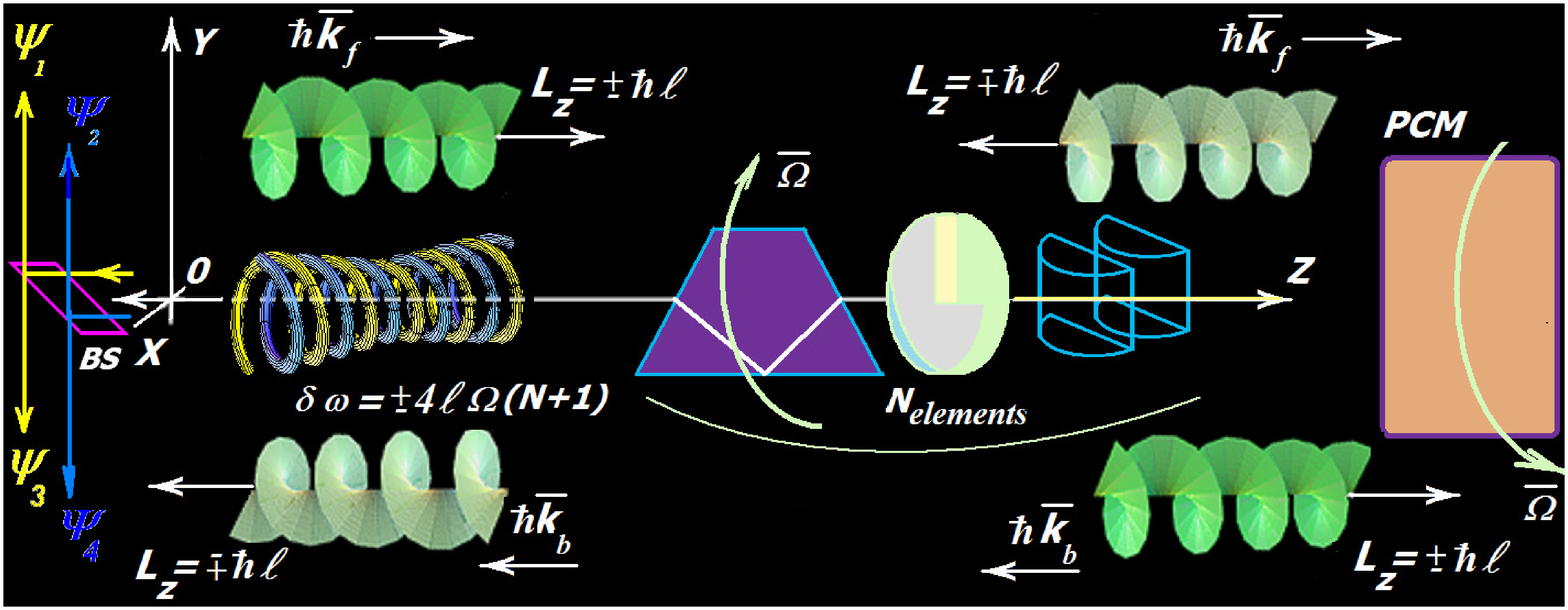}}
\caption{(Color online)Additivity of RDS in
PCVI 
inside rotating vehicle. PC-mirror counter rotating around 
$Z$-axis changes the carrier frequency of 
reflected photon 
to $\delta \omega=\pm 2 \ell \cdot  \Omega$. 
The sign of
$\delta \omega$ is positive when optical torque $\vec T$
produced by a bunch of the rotating photons upon mirror 
has the opposite direction 
compared to the rotation frequency of PCM $\vec \Omega$. 
When Dove prism rotates in opposite direction vs PCM 
the net RDS reaches six-fold value 
$\delta \omega=\pm 6 \ell \cdot  \Omega$ due to the additional 
OAM alternation. The sequence of $N$ counter rotating OAM altering 
elements (including helical waveplates and cylindrical lenses) 
will produce 
net RDS of $\delta \omega=4\ell\cdot   \Omega (N+1/2)$ value.
$|\Psi_{1,2,3,4}>$ designates antibunching 
of photons, which 
belongs to the $two$ (for $\ell=1$) 
helical wavefunctions, separated by $\lambda/2$ interval, 
deflected by entrance beamsplitter 
BS \cite {Woerdemann:2009}.}
\label{fig.3}
\end{figure}

\begin{equation}
\label {freq_shift_photon_PC_rot} 
\delta \omega = \omega_b - \omega_f = \pm 
2 \ell \cdot \Omega + \frac{2 \ell \cdot \hbar}{(I_{zz})_{PCM}}.
\end{equation} 
The second term in the right-hand side of (\ref {freq_shift_photon_PC_rot}) 
is negligible for typical masses ($m \sim g$) and sizes ($r \sim cm$) of 
a prisms and mirrors 
$\hbar /I_{zz} \sim \hbar / (m \cdot r^{-2}) \cong 10^{-27} Hz$.
as in Beth's \cite {Beth:1936} and Dholakia's \cite {Dholakia:2002D} experiments 
for the interaction of circularily polarized photons with the  
macroscopic object (half-wavelength plate).

The angular speed of rotation of the interference pattern proves 
to be $\dot \theta = \delta \omega / 2 \ell= \Omega $ thus 
the pattern rotates synchronously with the reference frame. 
Consequently the sole PCM cannot detect frame rotation. 
The helical interference pattern outside PCM 
will be dragged by helical diffraction grating 
\cite {Okulov:2008} within the phase-conjugating 
mirror. No atomic coherence \cite {Boyd:2011} 
is required in our case. 

Nevertheless there exists a possibility to accumulate the 
RDS by means of a chain of OAM alternating elements.   
To achieve the accumulation 
of RDS the adjacent 
components of PCVI 
must rotate in opposite 
directions $\vec \Omega_n=(-1)^n \vec \Omega$, where $n=0$ stands 
for PCM, $n=1$ for the adjacent Dove prism to PCM, $n=N$ for 
the last Dove prism(DP) near BS. This is necessary because 
OAM is altered 
after the passage of the Dove prism and the mutual orientation of 
the angular momenta of the photon and the next prism should 
be maintained throughout the chain. 
When $even$ elements (the PCM itself 
and N/2 Dove prisms) 
of PCVI are fixed in $\vec \Omega$ rotating 
frame and the rest N/2 $odd$ elements 
ought to rotate there with angular velocity $- 2 \vec \Omega$. 
The chain of the $N$ rotating OAM-alternating 
elements will produce the 
net rotational Doppler shift amounting to 
$\delta \omega=4\ell\cdot \Omega (N+1/2)$. 
Thus PCVI interference pattern (fig.\ref{fig.3}) 
will revolve with enhanced 
angular speed $\dot \theta$ of the frame rotation by the 
factor $\dot \theta = \pm 2 \cdot (N+1/2) \cdot \Omega $. 

For example PCVI (fig.\ref{fig.3}) may be used for 
demonstration of the possibility of detection 
of the sub-Hertzian 
rotation of the reference frame with the 
Earth $(\Omega_{\oplus} \sim 2\pi/{86400})$.  
The helical interference pattern 
will rotate much faster than Earth itself. 
Namely the equation $4 \ell \cdot (N+1/2)=24$ 
\cite {Dholakia:2002D} have the only one solution 
for integer $\ell, N$ ($\ell=4,N=1$). Hence 
the optical vortex with charge $\ell=4$ 
passed through single Dove prism rotating 
with $\vec \Omega_{\oplus}$ and PCM rotating in opposite 
direction $-\vec \Omega_{\oplus}$ will produce 
the $2 \cdot \ell$ 
spots of interference pattern. The reflection from 
entrance beamsplitter BS will cause one pass per hour
of the spot of interference pattern across the detector 
window, despite the Earth 
rotates once in 24 hours only. 

The further accumulation of RDS in PCVI 
might be achieved due to 
installing the $N=60$ counter rotating 
image-inverting elements and $\ell=6$ optical vortex. 
In such configuration the  
$2 \cdot \ell$ helices of interference pattern 
eq.(\ref {Archimedean_inter_patt}) 
will produce $2 \ell$  spots at the 
PCVI output (entrance beamsplitter BS) with 
a one pass through detector window within approximately 
each 60 seconds. 
For this purpose the even components (PCM and N/2 Dove prisms)
may be fixed at setup rotating with velocity $\vec \Omega_{\oplus}$
while the others N/2 prism should rotate 
with "bias" angular velocity $-2 \vec \Omega_{\oplus}$ with 
respect to rotating setup (rotating table).
This enhancement will model 
the detection Earth rotation, and this will alter 
the $ \ell \hbar$ OAM  
of the $each$ photon 121 times during one passage 
through PCVI. Noteworthy the Dove prism 
is not the sole element capable to alter the 
photon's OAM. This can be done as 
well with helical waveplates and cylindrical lenses 
\cite {Volostnikov:1989,Allen:1992}.
The helical interference pattern within PCVI might also 
be written by means of atomic coherence effects in a solid-state 
resonant medium \cite {Boyd:2011}. 

\section {Single-photon operation of the phase-conjugating 
vortex interferometer.}
The single-photon operation \cite{Kapon:2004} is based upon the 
superposition of the forward and backward quantum 
states with $\ell \hbar$ OAM: 
\begin{equation}
\label {wavefunction} 
 |\Psi>_{helix} = {\frac {1}{\sqrt {2}}}
( |\Psi_{\pm \ell \hbar}>_f+|\Psi_{\mp \ell \hbar}>_b)= 
{\frac {1}{\sqrt {2 \ell}}}{\sum_{j_{h}}}{|\Psi_{j_{h}}>}.
\end{equation}
The detection of this superposition is not a 
trivial two-detector 
procedure, because the interference pattern is composed of 
$2\ell$ twisted helices $|\Psi_{j_{h}}>$. 
The entrance beamsplitter BS will reflect both upward and downward 
the interference pattern \cite{Woerdemann:2009,Okulov:2008} 
composed of the $2 \ell$ spots 
located on an ellipse, rather than independent 
forward $|\Psi_{\pm \ell \hbar}>_f$ and 
backward $|\Psi_{\mp \ell \hbar}>_b$ photon states. 
For the simplest case $\ell=1$ the 
photon will be in the superposition state of the 
two $helical$  wavefunctions 
designated by appropriate colors at fig.(\ref {fig.3}): 
\begin{equation}
\label {wavefunction_helix} 
 |\Psi>_{helix} = {\frac {1}{\sqrt {2}}}( |\Psi_{Blue}> +|\Psi_{Yellow}>).
\end{equation}
This means that two 
detectors (for $|\Psi_{1}>$ and $|\Psi_{2}>$) placed above the entrance 
beamsplitter BS \cite{Woerdemann:2009} and 
two detectors located below BS 
(for $|\Psi_{3}>$ and $|\Psi_{4}>$) can indicate the 
$antibunching$ of the photons \cite{Scully:1997}, 
belonging to either of the two 
helices composing the interference pattern. As in a double-slit 
Young interference experiment the crude attempt of the 
eavesdropping the $which$ $way$ 
 photon moves (the forward or backward one) will 
destroy the helical interference pattern. 
On the other hand when single-photon 
quantum state is prepared as a 
toroidal pattern (fig.\ref{fig.2}) 
the photon belongs to the sequence of the 
equidistantly spaced 
toroidal Wannier wavefunctions $|\Psi_{j_{tor}}>$
separated by $\lambda /2$ intervals:    
\begin{equation}
\label {wavefunction_toroidal} 
|\Psi>_{tor} =
{\frac {1}{\sqrt {2}}}
 ( |\Psi_{\pm \ell \hbar}>_f+|\Psi_{\pm \ell \hbar}>_b)=
{\frac {1}{\sqrt {N_{tor}}}}{\sum_{j_{tor}}}{ \Psi_{j_{tor}}}.
\end{equation}
\section {Conclusion.}
	In summary we analyzed the phase conjugating vortex interferometer 
for the both single photon \cite{Kapon:2004} and the $cw$ 
laser output. Because of the alignment of the all optical components
along photon $Z$ propagation axis  PCVI looks promising 
from the point of view of rotation sensing \cite {Scully:1997}.

In PCVI the RDS $\delta \omega$ enhances 
the noninertial frame rotation $\vec \Omega$ by a factor of 
the even multiple of the photon's topological charge $\ell$ and 
of the number of angular momentum inverting elements $N$ in 
PCVI chain. 
Noteworthy that in the proposed measurement of the Earth rotation 
$\delta \omega=\pm 4\ell\cdot (N+1/2) \Omega_{\oplus} \cos(\phi)$ 
will show dependence on geographical latitude $\phi$ as 
it known for the Foucault pendulum \cite{Foucault:1852}: 
on the poles $\delta \omega$ will be equal to 
the maximum value 
when the angle $\phi$ between normal and PCVI axis is $0$ or $\pi$,
while at equator $\delta \omega$ might reach maximum value 
when PCVI axis is parallel to the Earth rotation axis. 
The preliminary analysis 
have shown that the laser linewidth of the order $\Delta \omega / 2 \pi \sim 10^3 Hz$ 
might be sufficient for Earth rotation detection by PCVI (fig.\ref{fig.3}). 
We hope to consider the above issues including $entanglement$ 
of the helical photons in PCVI due to mixing counter propagating 
photon vortex states via entrance beamspliter in a more details in the subsequent work \cite {Haus:1990}.

\end{document}